\documentclass[twocolumn,english,prb,aps,superscriptaddress]{revtex4}
\usepackage{amsmath}
\usepackage{amssymb}
\usepackage{graphicx}
\usepackage[USenglish]{babel}

\newcommand{\beq}{\begin{equation}}
\newcommand{\eeq}{\end{equation}}
\renewcommand{\(}{\left(}
\renewcommand{\)}{\right)}
\renewcommand{\[}{\left[}
\renewcommand{\]}{\right]}

\begin{document}

\title{Gap structure in Fe-based superconductors with accidental nodes: the role of hybridization}

\author{Alberto Hinojosa}
\affiliation{Department of Physics, University of Minnesota,
Minneapolis, Minnesota 55455, USA}
\author{Andrey V. Chubukov}
\affiliation{Department of Physics, University of Minnesota,
Minneapolis, Minnesota 55455, USA}

\begin{abstract}
We study the effects of hybridization between the two electron pockets in Fe-based superconductors with $s$-wave gap with accidental nodes. We argue that hybridization reconstructs the Fermi surfaces and also induces an additional inter-pocket pairing component. We analyze how these two effects modify the gap structure by tracing the position of the nodal points of the energy dispersions in the superconducting state. We find three possible outcomes. In the first, the nodes simply shift their positions in the Brilluoin zone; in the second, the nodes merge and disappear, in which case the gap function has either equal or opposite signs on the electron pockets; in the third, a new set of nodal points emerges, doubling the original number of nodes.
\end{abstract}

\maketitle

\section{Introduction}

The iron pnictides and chalcogens have been the subject of intense study since 2008, when it was discovered that they are superconducting at relatively high critical temperatures~\cite{hosono}.
 Understanding their gap structure is an important step toward identifying the mechanism responsible for superconductivity in these materials. Although the multi-band nature of Fe pnictides/chalcogenides allows for many different gap structures, the  prevailing scenario is that the pairing occurs between electrons on the same Fermi surface (FS) and the superconducting gap function has $s^{+-}$ symmetry, i.e., the gap  changes sign between hole and electron pockets.
 There is experimental evidence that in some members of the family, like BaFe$_2$(As$_{1-x}$P$_x$)$_2$,\cite{BaFeAsP} LaOFeP,\cite{LaOFeP} and LiFeP,\cite{LiFeP} the gap has nodes, likely on the electron pockets.

\begin{figure}[htb]
	\centering
		\includegraphics[width=0.3\textwidth]{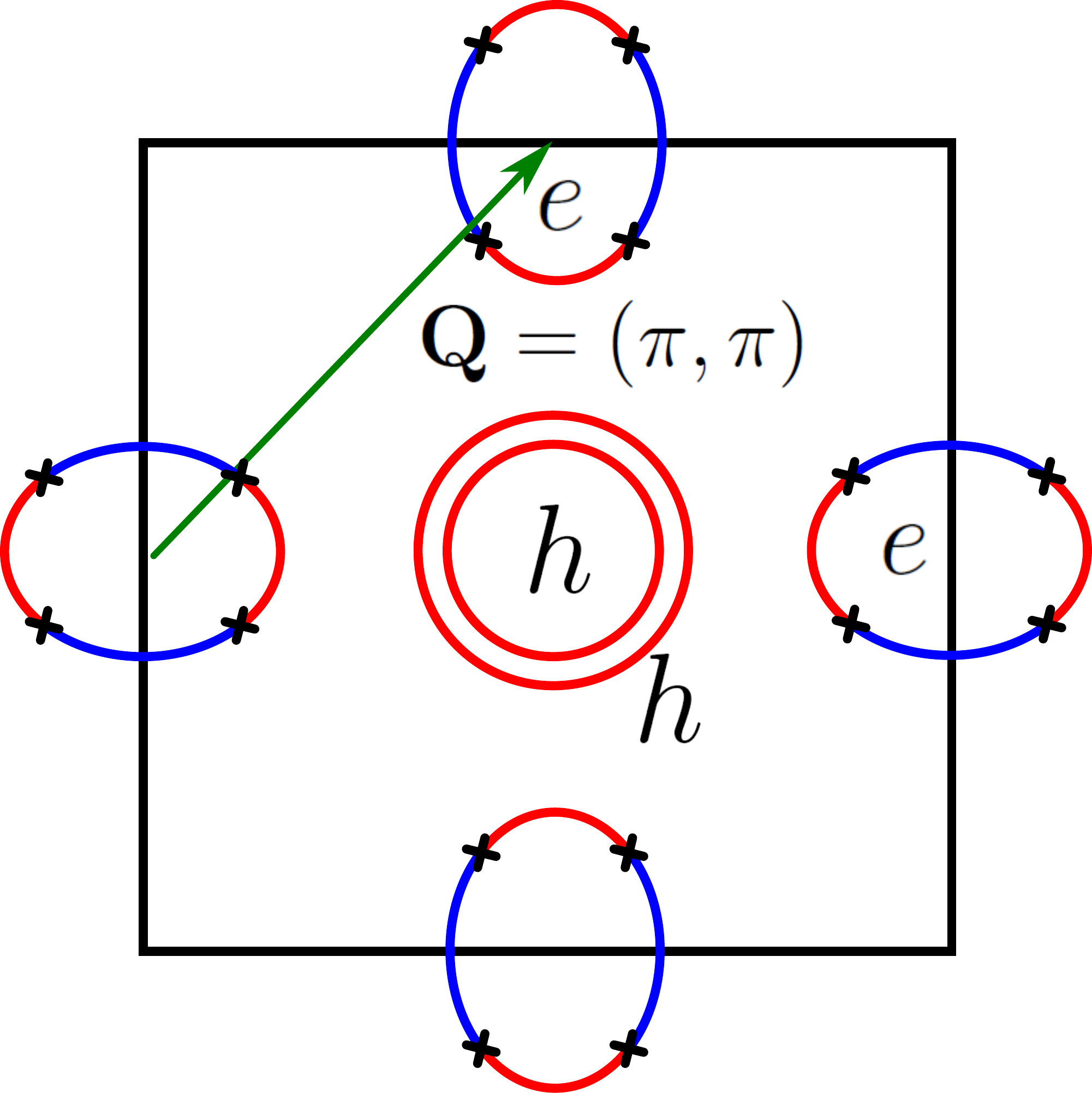}
	\caption{Unfolded Brillouin zone with one Fe atom per unit cell (with no hybridization). Holes and electron pockets are labeled by $h$ and $e$, respectively. The crosses on the electron pockets indicate where the nodal lines of the gap function intersect the Fermi surface.}%AH: Modified caption.
	\label{fig:multipocket_nodes}
\end{figure}

Previous studies of the gap structure were mostly restricted to an Fe-only approach, in which a generic model of the band structure consists of two nearly circular hole pockets centered at $(0,0)$ and two elliptical electron pockets centered at $(\pi,0)$ and $(0,\pi)$ in the first Brillouin zone (BZ) (see Figure \ref{fig:multipocket_nodes}). In some systems there exists, at least for some $k_z$, a third hole pocket, centered at $(\pi,\pi)$.
%AC
The $s^{+-}$ superconductivity is believed to be chiefly caused by a magnetically enhanced interaction between hole and electron pockets~\cite{review}.
 The nodes on the two electron pockets come about because by symmetry the $s$-wave gap on these pockets has the form $\Delta (1 \pm \alpha\cos 2 \theta_k)$
 (plus higher harmonics), and if $\alpha >1$, the gap vanishes at $\cos 2 \theta_k = \pm 1/\alpha$.~\cite{Kemper2010,vavilov,Maiti2011}.

However, this Fe-only scenario is incomplete because the electron hopping between Fe atoms predominantly occurs via pnictogen or chalcogen sites, half of which
  are located above and half below each Fe layer in a checkerboard pattern~\cite{hybrid}. As a result, the actual symmetry is lower than that of the Fe-only lattice, and the correct unit cell contains two Fe atoms.  The non-equivalence of hopping from above and from below an iron layer causes the electron pockets to hybridize. In this paper we will study the effect of this hybridization on the gap structure.

  In order to incorporate this effect into models with one Fe atoms, one has to include additional terms in the Hamiltonian with excess momentum $\mathbf{Q}=(\pi,\pi)$. This does not actually violate conservation of momentum because this vector folds into a reciprocal lattice vector in the actual BZ with two Fe atoms per unit cell.  The momentum $\mathbf{Q}$ connects the two electron FSs and also the hole pocket centered at $(\pi,\pi)$ with the other two hole pockets. Our primary goal will be to study how the accidental nodes on the electron pockets evolve once we include hybridization. Therefore, we focus on the effect of hybridization on the electron pockets.

  The hybridization gives rise to two effects.   First, hopping via a pnictogen/chalcogen sites gives rise to an additional quadratic term in the Hamiltonian for two electron pockets
\beq\label{eq:Hlambda}
\mathcal{H}_\lambda=\sum_\mathbf{k}\[ \lambda_k c^\dagger_{\mathbf{k}\alpha} d_{\mathbf{k+Q}\alpha}+ \lambda^*_k d^\dagger_{\mathbf{k}\alpha} c_{\mathbf{k+Q}\alpha}\],
\eeq
where $c$ and $d$ are operators for electrons near each of the two electron FSs (we discuss the form of $\lambda_k$ in the next section)
 and the sum over repeated spin indices is implied.
 This cross-term mixes the two electron pockets and reconstructs the electron FSs.
 Second, there appear new four-fermion interaction terms in which incoming and outgoing momenta differ by $\mathbf{Q}$. %AH: changed $(\pi,\pi)$ to Q.
  In the superconducting state, in which we are  interested, two out of four fermions can be put into the condensate and the four-fermion terms with excess momentum $\mathbf{Q}$
	reduce to  quadratic terms with prefactors proportional to the superconducting gap. These new terms describe inter-pocket pairing between fermions from two different electron pockets:
 \begin{equation}
	\mathcal{H}_\beta =\frac{1}{2}\sum_{\mathbf{k}} \beta_k \left[c^\dagger_{\mathbf{k}\alpha}d^\dagger_{-\mathbf{k}-Q\beta}+d^\dagger_{\mathbf{k}\alpha}c^\dagger_{-\mathbf{k}-Q\beta}\right]i\sigma^y_{\alpha\beta}+\mathrm{H.c.}
	\end{equation}
In other words, due to hybridization, the non-zero intra-pocket pairing condensates   $\langle c^\dagger_{\mathbf{k}\alpha}c^\dagger_{-\mathbf{k}\beta}\rangle$ and $\langle d^\dagger_{\mathbf{k+Q}\alpha}d^\dagger_{-\mathbf{k-Q}\beta}\rangle$ induce inter-pocket pairing between the two electron pockets.

In this paper we study how the additional terms $\mathcal{H}_\lambda$ and $\mathcal{H}_\beta$  affect the gap structure when nodes are present on the electron pockets. The effect of the hopping $\lambda_\mathbf{k}$ term alone has been studied before\cite{Khodas2012b}, but not its interplay with the pairing term. We find that the hopping and inter-pocket pairing terms generally pull the nodal points in opposite direction. If the $\lambda_\mathbf{k}$ term dominates and reaches a certain threshold value, the nodes merge and disappear at particular symmetry points, and the gap  acquires a uniform and equal phase on the two electron pockets (opposite to the phase on the hole pockets). In contrast, when $\beta_k$ dominates and  reaches a threshold value, the nodes merge and disappear at a different set of symmetry points and the phase of the superconducting order parameter becomes  opposite on the hybridized electron pockets. This is the same gap structure
  that was recently found in the analysis of pairing in the orbital formalism\cite{Kotliar2014} and dubbed orbital anti-phase $s^{+-}$ state.
   The state  with opposite signs of the gaps on the two electron pockets has also been found in the analysis of possible superconducting states in LiFeAs, albeit for a different reason~\cite{ilya_li}.
     As an interesting peculiarity, we found that for elliptical pockets nodes disappear in a rather non-trivial way --
      first new nodes appear and the number of nodes doubles, and then the new and already existing nodes merge and disappear.
         Such behavior has not been found before in the studies of multi-band superconductors, as far as  we know.

\begin{figure}[htb]
	\centering
		\includegraphics[width=0.5\textwidth]{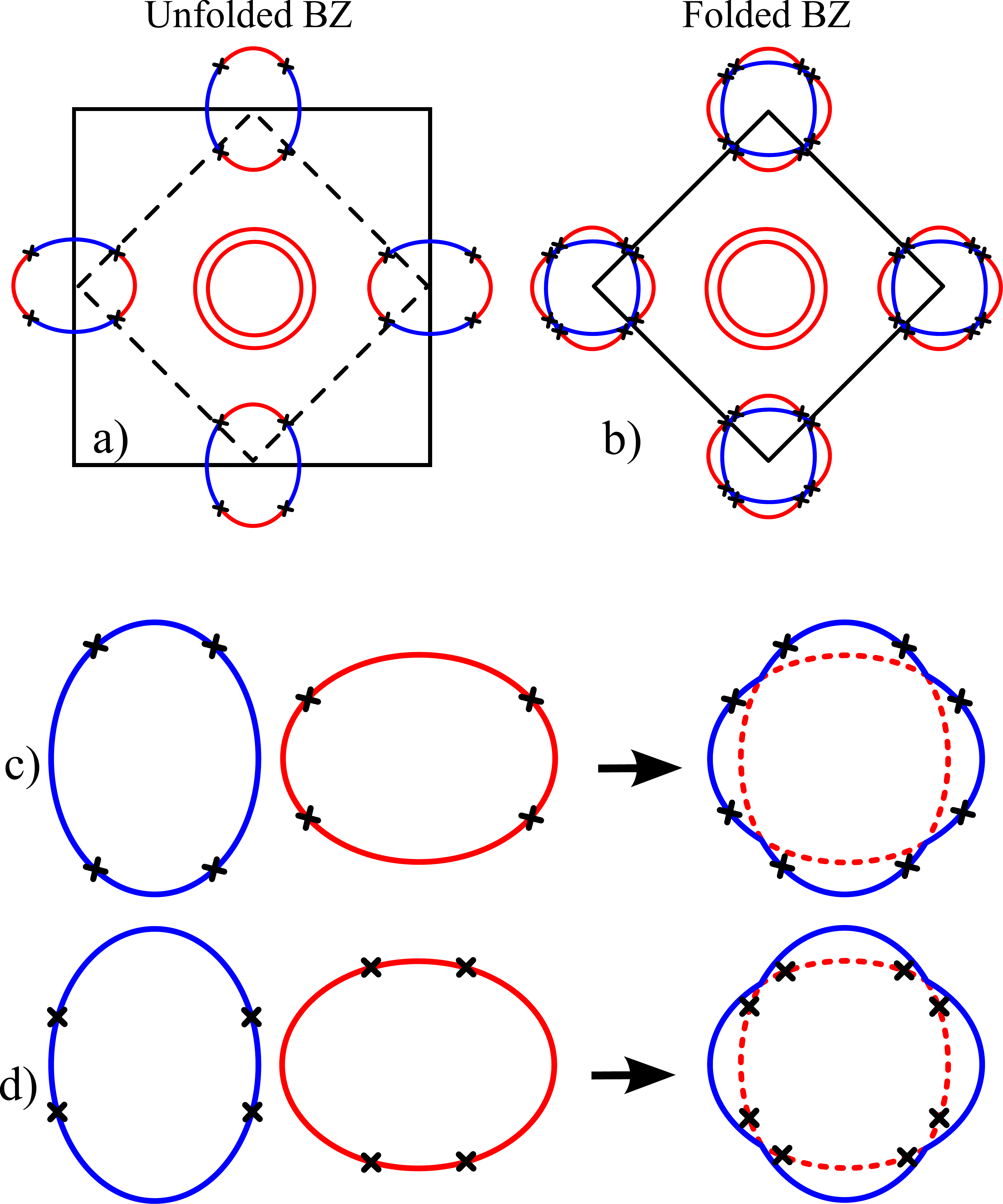}
	\caption{Fermi surface folding. a) Unfolded Brillouin zone corresponding to a 1Fe unit cell. b) Folded Brillouin zone corresponding to a 2Fe unit cell.  The electron Fermi surfaces overlap and are reconstructed into outer and inner parts. Depending on their original location, the nodes of the gap function are either all located on the outer surface, as shown in c), or on the inner surface, as shown in d).}
	\label{fig:folding}
\end{figure}

The hybridization between electron pockets, either due to $\mathcal{H}_\lambda$ or to $\mathcal{H}_\beta$, has to be distinguished from the
effect of the folding of the 1Fe BZ into the 2Fe BZ.  Upon folding, fermionic momenta transform as  ${\tilde k}_x = k_x + k_y, {\tilde k}_y = k_y-k_x$,
  and the two electron FSs, originally centered at $(0,\pi)$ and $(\pi,0)$, merge around $(\pi,\pi)$ (see Fig. \ref{fig:folding}).  The inner and outer FSs touch each other along ${\tilde k}_x = \pi$ or ${\tilde k}_y = \pi$. The merging can be viewed as ``reconstruction'' of the two electron FSs into
   an inner FS with no nodes and an outer FS with 8 nodes, or vice versa.  However, this ``reconstruction" is just a rotation in momentum space and a re-branding.
   The $k_F$ remains the same and the location of the nodes remains at exactly the same
	angles as without
folding, only the reference axis rotates by $45^o$.   The hybridization is a different phenomenon -- it actually reconstructs the original FSs into
 an inner and an outer FSs at {\it new} $k_F$  and creates a pairing component between them. As a consequence, the position of the nodal points shifts.

The paper is organized as follows: In Section \ref{sec:model} we discuss our model. In Section \ref{sec:circular}, we consider, as a warm-up,
 the limiting case of circular electron pockets and analyze the two effects of hybridization first separately and then together and study their interplay. Section \ref{sec:elliptical} extends the analysis to the more general case of elliptical pockets. We summarize our findings in Section \ref{sec:conclusions}.

\section{The model}\label{sec:model}

 We consider a two-dimensional multi-band model with hole pockets centered at $(0,0)$ and elliptical electron pockets centered at $(0,\pi)$ and $(\pi,0)$ in the unfolded BZ.  In the normal state, the Hamiltonian describing the two non-hybridized electron pockets is simply
\begin{equation}
	\mathcal{H}_0 =\sum_{\mathbf{k}} \left[\xi^c_\mathbf{k}c^\dagger_{\mathbf{k}\alpha}c_{\mathbf{k}\alpha}+\xi^d_\mathbf{k}d^\dagger_{\mathbf{k}\alpha}d_{\mathbf{k}\alpha}\right].
\end{equation}

We assume that the dominant interaction which leads to $s^{+-}$ superconductivity
   is the repulsion between electron and hole pockets, enhanced by $(\pi,\pi)$ spin fluctuations (in the 2Fe BZ).
      The s-wave gap on the hole pockets is invariant under rotations by $\pi/2$, so it can be expanded in $\cos 4n\phi$ harmonics, where $\phi$ is the angle along the hole pocket and $n$ is an integer. On the electron pockets the expansion of the $s-$wave gap contains $\cos 2n\theta$ terms, where $\theta$ is the angle along the electron pockets and the components with odd $n = 1,3,\ldots$ have opposite signs on the two electron pockets, if we measure $\theta$ from the same direction for both pockets~\cite{vavilov}.
        These odd multiples  of $2\theta$ are allowed because the electron pockets transform into each other under a rotation by $\pi/2$.  Numerical analysis\cite{Kemper2010, Maiti2011} shows that the gap on the electron pockets can be well-approximated
         by the two first harmonics $n=0$ and $n=1$, whose magnitudes are generally comparable to each other.
          Accordingly, we set $\Delta_{e1} (\theta_\mathbf{k})= \Delta (1- \alpha\cos 2\theta_\mathbf{k})$, $\Delta_{e2} (\theta_\mathbf{k+Q})= \Delta (1+ \alpha\cos 2\theta_\mathbf{k+Q})$.  The corresponding term in the Hamiltonian is thus
\begin{align}
	\mathcal{H}_\Delta &=\frac{1}{2}\sum_{\mathbf{k}} \Delta \left[(1-y_\mathbf{k})c^\dagger_{\mathbf{k}\alpha}c^\dagger_{-\mathbf{k}\beta}\right.\\
	&\quad\quad\quad\quad\quad+ \left.(1+y_\mathbf{k+Q})d^\dagger_{\mathbf{k+Q}\alpha}d^\dagger_{-\mathbf{k-Q}\beta}\right]i\sigma^y_{\alpha\beta}+\mathrm{H.c.},\nonumber
\end{align}
where $y_\mathbf{k}\equiv \alpha \cos 2\theta_\mathbf{k}$ and $y_\mathbf{k+Q}\equiv \alpha \cos 2\theta_\mathbf{k+Q}$.

Our goal is to analyze how accidental nodes on electron pockets evolve with increasing hybridization.   For this we assume from the beginning that $|\alpha| >1$ in which case the gaps on electron pockets have nodes when $y_\mathbf{k} = 1$ (for $c-$fermions) and $y_\mathbf{k+Q} =-1$ for $d-$fermions.

To simplify the presentation, we fold the 1Fe BZ into the 2Fe BZ and replace the momentum ${\bf k} + {\bf Q}$ of d-fermions  by ${\bf k}$.  The momenta ${\bf k}$ below are defined as a deviation from ${\bf Q}$, which is the location of the electron pockets in the folded BZ.
%AC
In order to preserve the $\cos 2\theta_\mathbf{k}$ form of the gap function, we  define $\theta_\mathbf{k}$ relative to the minor axis of the $c$ pocket.

In the normal state, the inclusion of hopping via pnictogen/chalcogen atoms generates  mixing between $c$ and $d$ fermions:
\begin{align}
	\mathcal{H}_\lambda &=\sum_\mathbf{k}\[ \lambda_k c^\dagger_{\mathbf{k}\alpha} d_{\mathbf{k}\alpha}+ \lambda^*_k d^\dagger_{\mathbf{k}\alpha} c_{\mathbf{k}\alpha}\],
\end{align}

A microscopic derivation  of $\lambda_\mathbf{k}$ shows~\cite{oskar,mazin_i,sadovskii,Khodas2012b}  that in 1111 systems (in which the configuration of pnictide atoms around every Fe layer is the same), $\lambda_\mathbf{k}$ vanishes along the lines $k_x=\pm k_y$ and has some weak $k_z$ dependence. In 122 structures (in which the ``above/below" configuration of pnictogen/chalcogen atoms is inverted from one Fe layer to the other), $\lambda_\mathbf{k}$ has minima but does not vanish along any direction. In the presence of a spin-orbit interaction $\lambda_\mathbf{k}$  does not have zeros even in 1111 systems~\cite{oskar,mazin_i}.  Because our goal is to understand the generic effect of the hybridization between the two bands, we will treat $\lambda_\mathbf{k}$ as a constant $\lambda$ \cite{mazin_i,Coldea}. Earlier analysis of the effect of $\lambda_\mathbf{k}$ including its angular dependence (but without interplay with inter-pocket pairing) has shown that the results are qualitatively the same as for constant $\lambda$\cite{Khodas2012b}.

   We next consider how hybridization affects the  pairing terms.  They can be subdivided into two types.
Terms of the first type  describe an interaction with excess momentum $\mathbf{Q}$ between electron pockets and contain three fermionic operators from one pocket and one from the other pocket\cite{Khodas2012a}, e.g.,
 \begin{align}\label{eq:HQ1}
	\mathcal{H}_1= u_1 \sum_{\mathbf{k},\mathbf{p},\mathbf{q}}&\bigg[\(c^\dagger_{\mathbf{k}\alpha} d_{\mathbf{k}-\mathbf{q},\alpha}+d^\dagger_{\mathbf{k}\alpha} c_{\mathbf{k}-\mathbf{q},\alpha}\) \nonumber\\
	&\times \(c^\dagger_{\mathbf{p}\beta} c_{\mathbf{p}+\mathbf{q},\beta}+d^\dagger_{\mathbf{p}\beta} d_{\mathbf{p}+\mathbf{q},\beta}\)\bigg].
\end{align}
Terms of the second type  contain an interaction with excess momentum $\mathbf{Q}$ involving two fermions from a hole pocket and two from different electron pockets, e.g.,
\begin{align}\label{eq:HQ2}
	\mathcal{H}_2= u_2 \sum_{\mathbf{k},\mathbf{p},\mathbf{q}}&\bigg[\(c^\dagger_{\mathbf{k}\alpha} h_{\mathbf{k}-\mathbf{q},\alpha}+h^\dagger_{\mathbf{k}\alpha} c_{\mathbf{k}-\mathbf{q},\alpha}\) \nonumber\\
	&\times \(d^\dagger_{\mathbf{p}\beta} h_{\mathbf{p}+\mathbf{q},\beta}+h^\dagger_{\mathbf{p}\beta} d_{\mathbf{p}+\mathbf{q},\beta}\)\bigg],
\end{align}
where the operator $h_{\mathbf{p}\alpha}$ describes fermions near one of the hole pockets.  The two types of terms are different,  yet
  their effect on the $s^{+-}$ superconducting  state is the same -- both induce an additional pairing interaction between fermions belonging to different electron pockets. Indeed, in the superconducting state $\langle c^\dagger_{\mathbf{k}\alpha}c^\dagger_{-\mathbf{k}\beta}\rangle$, $\langle d^\dagger_{\mathbf{k}\alpha}d^\dagger_{-\mathbf{k}\beta}\rangle$, and $\langle h^\dagger_{\mathbf{k}\alpha}h^\dagger_{-\mathbf{k}\beta}\rangle$ are all non-zero. Decoupling  four-fermion terms in (\ref{eq:HQ1}) and (\ref{eq:HQ2}) using these averages, we obtain anomalous quadratic terms involving $c$ and $d$ fermions:
 \begin{equation}
	\mathcal{H}_\beta =\frac{1}{2}\sum_{\mathbf{k}} \beta_k \left[c^\dagger_{\mathbf{k}\alpha}d^\dagger_{-\mathbf{k}\beta}+d^\dagger_{\mathbf{k}\alpha}
c^\dagger_{-\mathbf{k}\beta}\right]i\sigma^y_{\alpha\beta}+\mathrm{H.c.}.	\end{equation}
The coupling $\beta_k$ is proportional to the magnitude of the $s^{+-}$ gap $\Delta$ and has some non-singular  angular dependence  which we can safely neglect.

We assume without loss of generality that the parameters $\Delta$ and $\lambda$ are real and positive. The parameter $\beta \propto \Delta$ is then also real, but its sign can be either positive or negative.

 Below we consider various ratios of $\lambda/\beta$  and two  FS geometries. In each case we compute the  quasi-particle dispersion in the superconducting state and determine the position of the nodal points.  In all cases we find two different dispersions: One is gapped over the entire BZ, while the other contains nodal points in a subset of the parameter space.

\section{Circular pockets}\label{sec:circular}

As a warm-up, consider the limiting case when the two electron pockets are identical and have full rotational symmetry, i.e., $\xi^c_\mathbf{k}=\xi^f_\mathbf{k}\equiv \xi_\mathbf{k}$.

\subsection{Inter-pocket pairing only ($\beta\neq 0$, $\lambda= 0$)}

\begin{figure*}[htb]
	\centering
		\includegraphics[width=0.8\textwidth]{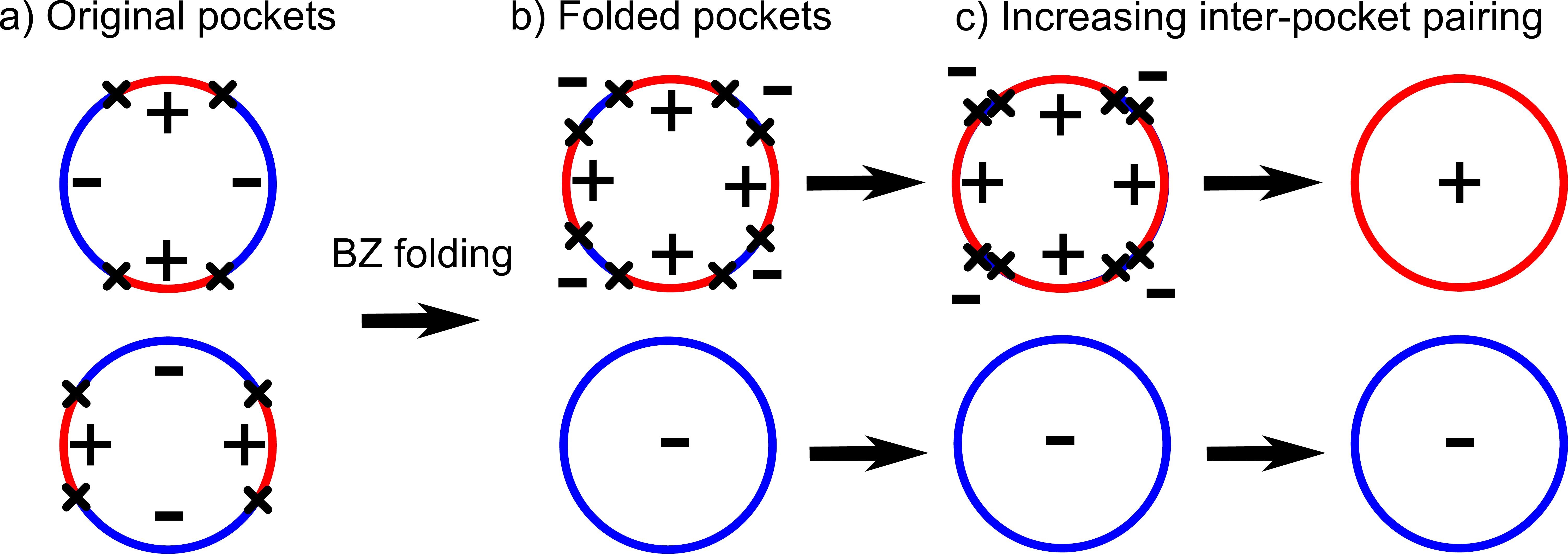}
	\caption{Gap structure for circular electron pockets with inter-pocket pairing. The crosses represent the location of the nodal points of the quasi-particle dispersions in the superconducting state. In a) and b) we show the unfolded and folded zones, respectively. In the folded zone, the pockets should overlap but we separate them for clarity. As we increase the inter-pocket pairing, keeping the conventional intra-pocket pairing fixed, the nodal points shift toward the diagonal lines $k_x=\pm k_y$, as seen in c). If the inter-pocket pairing reaches a critical value, the nodes merge and disappear, resulting in electron Fermi surfaces with opposite signs of the gap function.}
	\label{fig:circular_beta}
\end{figure*}

In this special case, we find, after straightforward diagonalization of the quadratic form,
 that the two dispersions in the superconducting state are
\begin{equation}
	\left(E_\mathbf{k}^\pm\right)^2 = \xi_\mathbf{k}^2+\(\Delta \pm \sqrt{\Delta^2 y_\mathbf{k}^2+\beta^2} \)^2,
\end{equation}
where the expression in parenthesis represents an effective gap function. At $\beta =0$, $E^+_\mathbf{k} = \pm \sqrt{\xi^2_\mathbf{k} + \Delta^2 (1 + |y_\mathbf{k}|)^2}$ and $E^-_\mathbf{k} = \pm \sqrt{\xi^2_\mathbf{k} + \Delta^2 (1 - |y_\mathbf{k}|)^2}$. This corresponds
 to the gap structure in the folded BZ:  one band has no nodes and the other band has 8 nodes (see Fig. \ref{fig:circular_beta}).

At $\beta \neq 0$, both dispersions evolve. The nodal points are still located on the FS (the locus of zero energy points in the normal state, given by $\xi_\mathbf{k}=0$).
 The band with energy $E^+_\mathbf{k}$ is shifted up at a non-zero $\beta$  and its effective gap function is definitely nodeless.
  In contrast, the band with energy $E^-_\mathbf{k}$ is shifted down and the positions of the 8 nodes shift to
\begin{equation}
	\cos(2\theta_\mathbf{k})=\pm\frac{\sqrt{\Delta^2-\beta^2}}{\alpha\Delta}.
\end{equation}
As $\beta$ increases, the nodal points move toward the diagonals $k_x=\pm k_y$, as shown in Fig. \ref{fig:circular_beta}. At a critical value $\beta_c=\Delta$ they meet along the BZ diagonals. At larger $\beta>\beta_c$ the nodes disappear.  At the same, because the nodes merge along the BZ diagonals, the sign of the gap on one FS is opposite to that on another FS. Such a gap structure has been obtained before in the analysis of possible gap configurations in LiFeAs both in orbital formalism and in band formalism~\cite{Kotliar2014,ilya_li}. In the orbital formalism, such a state was termed ``orbital antiphase'' $s^{+-}$  state\cite{Kotliar2014}.

\begin{figure}[htb]
	\centering
		\includegraphics[width=0.47\textwidth]{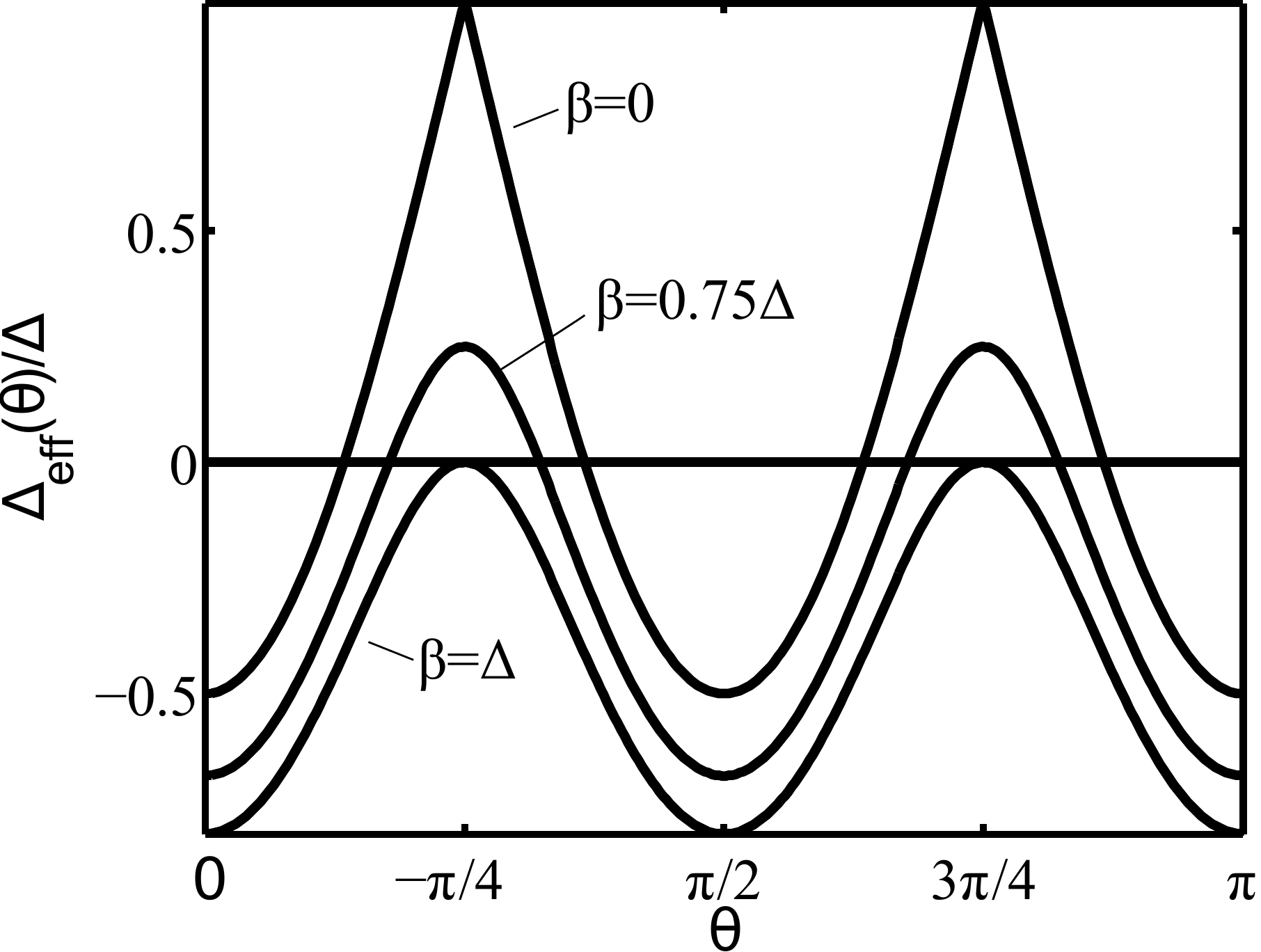}
	\caption{Nodal gap function with inter-pocket pairing $\beta$ evaluated over the FS. As $\beta$ increases, the function shifts downward and its zeroes move toward the angles $n\pi/2$, where $n$ is an integer, which correspond to the directions given by $k_x=\pm k_y$. At $\beta=\Delta$, pairs of zeroes merge at those angles. For $\beta>\Delta$, the function is negative and has no zeroes.}
	\label{fig:Delta_b}
\end{figure}

An intuitive way to understand this behavior is the following: the gap function at the upper band is positive, while the one  at the lower band has a roughly sinusoidal shape that crosses zero eight times, and its maxima occur at the diagonal directions $k_x=k_y$ and $k_x=-k_y$. As $|\beta|$ increases, the gap function shifts downward and thus the nodes  move towards the BZ diagonals, until $|\beta|$ reaches the critical value $\beta_c$. At this point, pairs of nodal points meet and annihilate. At larger
  $|\beta|>\Delta$ the maxima of the gap function are located below zero, i.e., the gap is negative for all angles. %AH: Changed 'roughly has a sinusoidal shape' to 'has a roughly sinusoidal shape'.
%AC need figure
This behavior is illustrated in Fig. \ref{fig:Delta_b}.

\subsection{Inter-pocket hopping only ($\lambda\neq 0$, $\beta=0$)}

\begin{figure*}[htb]
	\centering
		\includegraphics[width=0.8\textwidth]{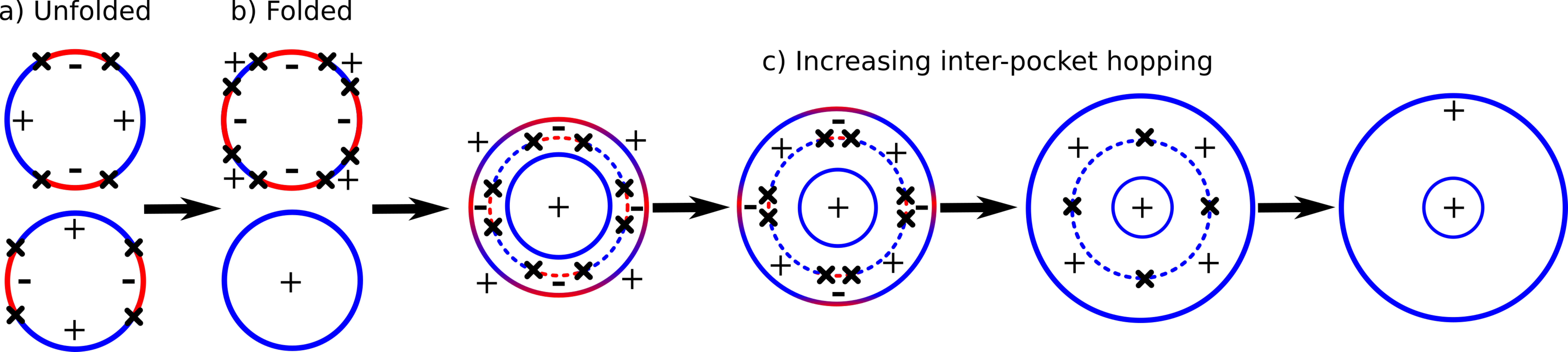}%AH: Slight modification in the caption because the figure now has a), b), and c).
	\caption{Gap structure for circular pockets with inter-pocket hopping. In a) and b) we show the unfolded and folded gap structure in the absence of hybridization, respectively. Inter-pocket hopping reconstructs the FSs as shown in c). The dashed line represents the original pockets, while the solid lines are the new set of two pockets. As the hopping parameter increases, the inner FS shrinks and the outer one becomes larger. In the superconducting state, the nodal points (represented as crosses) lie on the unhybridized FSs and shift toward the $k_x$ and $k_y$ axis as the hopping increases. At a critical value of this parameter the nodes meet and merge in pairs. For greater values of the parameter they vanish and the sign of the gap function becomes the same on both FSs. The separation between the FSs is purely schematic and has been exaggerated for clarity.}
	\label{fig:circular_lambda}
\end{figure*}

This case has been studied before,\cite{Khodas2012b} and we briefly summarize the results for completeness. The hybridization of the electron pockets
reconstructs the FSs. In order to study this effect we diagonalize the Hamiltonian $\mathcal{H}_0+\mathcal{H}_\lambda$ by introducing new quasi-particles $a$ and $b$  via
\beq
	c_{\mathbf{k}\alpha}=\frac{a_{\mathbf{k}\alpha}-b_{\mathbf{k}\alpha}}{\sqrt{2}},\quad
	d_{\mathbf{k}\alpha}=\frac{a_{\mathbf{k}\alpha}+b_{\mathbf{k}\alpha}}{\sqrt{2}}.\label{eq:transformation}
\eeq
After diagonalization, the non-pairing part of the Hamiltonian takes the form
\begin{equation}
	\mathcal{H}_0' =\sum_{\mathbf{k}} \left[\xi^a_\mathbf{k}a^\dagger_{\mathbf{k}\alpha}a_{\mathbf{k}\alpha}+\xi^b_\mathbf{k}b^\dagger_{\mathbf{k}\alpha}b_{\mathbf{k}\alpha}\right],
\end{equation}
where the dispersions are $\xi^a_\mathbf{k}=\xi_\mathbf{k}+\lambda$ and $\xi^b_\mathbf{k}=\xi_\mathbf{k}-\lambda$. The new FSs are concentric circles. The $a$ pocket is smaller and the $b$ pocket is larger than the unhybridized pockets.

In order to study the superconducting state, it is convenient to first rewrite $\mathcal{H}_\Delta$ in terms of the new operators $a$ and $b$ as
\begin{align}\label{eq:H_Delta_1}
	\mathcal{H}_{\Delta}' &=\frac{1}{2}\sum_{\mathbf{k}} \Delta\left[ a^\dagger_{\mathbf{k}\alpha}a^\dagger_{-\mathbf{k}\beta}+ b^\dagger_{\mathbf{k}\alpha}b^\dagger_{-\mathbf{k}\beta}\right]i\sigma^y_{\alpha\beta}\\
	&+\frac{1}{2}\sum_{\mathbf{k}} \Delta y_\mathbf{k} \left[a^\dagger_{\mathbf{k}\alpha}b^\dagger_{-\mathbf{k}\beta}+b^\dagger_{\mathbf{k}\alpha}a^\dagger_{-\mathbf{k}\beta}\right]\nonumber
i\sigma^y_{\alpha\beta}+\mathrm{H.c.}	
\end{align}

 Note that the inter-pocket pairing component $\Delta y_\mathbf{k}$ emerges. Diagonalizing the new Hamiltonian $\mathcal{H}_0'+\mathcal{H}_{\Delta}'$, we find the two dispersions for the quasi-particles in the superconducting state given by
\begin{equation}
	\left(E_\mathbf{k}^\pm\right)^2 = A_\mathbf{k}\pm\sqrt{B_\mathbf{k}},
\label{eq:H_Delta_2}
\end{equation}
where
\begin{align}
	A_\mathbf{k}=&\xi_\mathbf{k}^2+\Delta^2 (1+y_\mathbf{k}^2)+\lambda^2,\\
	B_\mathbf{k}=&4\left[ (\xi_\mathbf{k}\lambda)^2 +\Delta^2 y_\mathbf{k}^2 (\Delta^2+\lambda^2)\right],
\end{align}

%AC
 The dispersion $E^+_\mathbf{k}$, as defined in (\ref{eq:H_Delta_2}), is positive for all $\mathbf{k}$ even when evaluated at $\Delta=0$, so it has no locus of the nodal points. Both FS lines $\xi_\mathbf{k} = \pm \lambda$ in the normal state are part of the other dispersion $E^{-}_\mathbf{k}$. Once $\Delta$ becomes non-zero,
 one can easily check that $E^{-}_k$ is non-zero along the normal state FSs. However, on the original, non-reconstructed FS,
\begin{equation}
  E_\mathbf{k}^- = \sqrt{\Delta^2 + \lambda^2} - \Delta |y_\mathbf{k}|
\end{equation}
 This function contains 8 nodal points located at
\begin{equation}
	\cos(2\theta_\mathbf{k})=\pm\frac{\sqrt{\Delta^2+\lambda^2}}{\alpha\Delta}.
\end{equation}
As one increases $\lambda$, the nodes stay on the unhybridized FSs, %AH: changed 'original' to 'unhybridized'
 but move toward the $k_x$ or $k_y$ axes (whichever is closer), until $\lambda$ reaches a critical value $\lambda_c=\Delta \sqrt{\alpha^2-1}$. At this value of $\lambda$, pairs of nodal points merge  and then disappear at larger $\lambda$.
We show this schematically in Fig. \ref{fig:circular_lambda}.

The analysis of the signs of the gap requires some care.
  For $\lambda \gg \Delta$, the inter-pocket pairing term becomes irrelevant as the two reconstructed FSs are far apart from each other. In this limit, the gap on each reconstructed pocket is given by the fist line in (\ref{eq:H_Delta_1}) and is just $\Delta$ for both pockets.
In this limit,  the sign of the gap is indeed the same on both FSs.
At intermediate $\lambda$, however, one
    cannot define the phase of the gap function on the two FSs because in terms of the hybridized
  fermions the gap has contributions from both inter-pocket and inter-pocket condensates $\langle a_{\mathbf{k},\alpha} a_{-\mathbf{k},\beta} \rangle i \sigma^y_{\alpha \beta}$
  and $\langle a_{\mathbf{k},\alpha} b_{-\mathbf{k},\beta} \rangle i \sigma^y_{\alpha \beta}$, respectively. Because the limiting behavior at large $\lambda$ is known, it is ``natural'' to define both finite gaps with the same sign, as in Eq. (\ref{eq:H_Delta_1}),
   but we caution that this is rigorously justified only in the limit of very large $\lambda$.

\subsection{Inter-pocket pairing and hopping ($\beta\neq0$, $\lambda\neq 0$)}

\begin{figure*}[htb]
	\centering
		\includegraphics[width=0.8\textwidth]{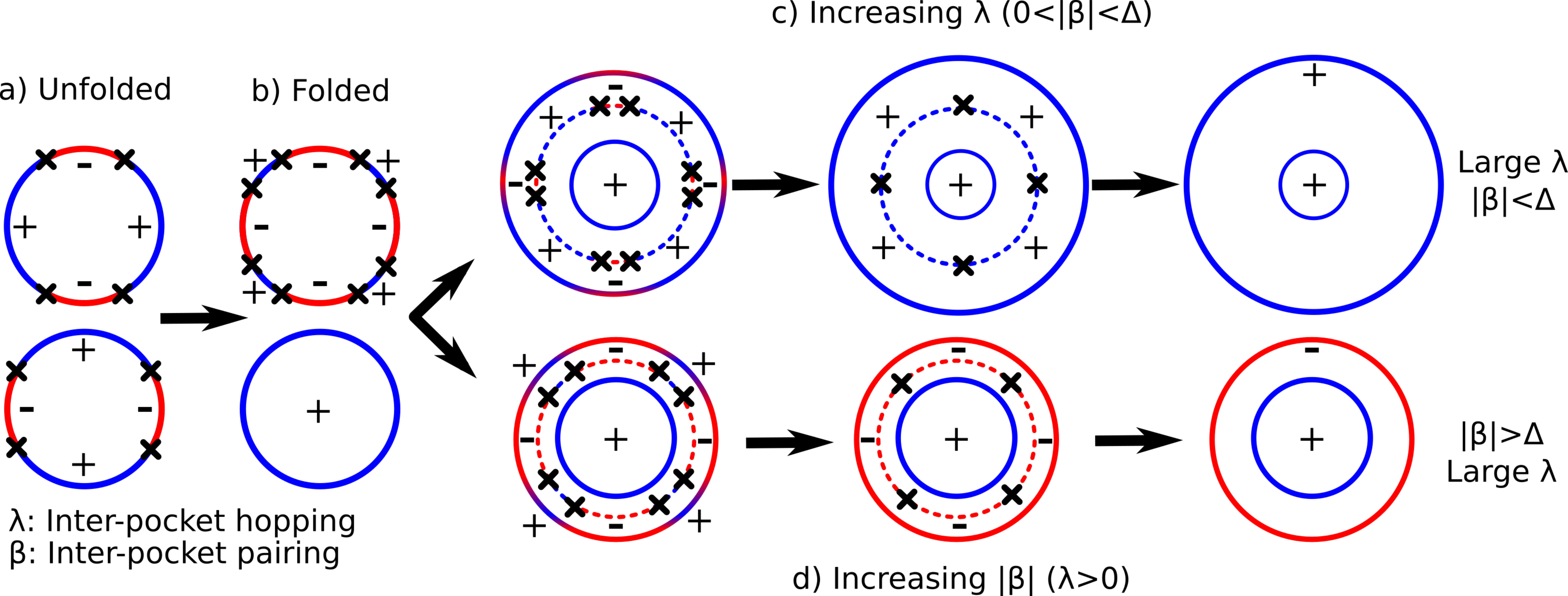}
	\caption{Gap structure for circular pockets with inter-pocket hopping and pairing interaction. The solid outer and inner circles represent the reconstructed FSs after hybridization. In the superconducting state, the nodal points (represented as crosses) lie on a circle (dashed line) but no longer on the unhybridized Fermi surface. As the hopping parameter $\lambda$ increases, the nodes move toward the $k_x$ and $k_y$ axis, where they can merge and disappear. In this scenario the gap function has the same sign on both electron pockets. Increasing the pairing strength $\beta$ shifts the nodes toward the diagonal lines $k_x=\pm k_y$ instead. The nodes can merge and vanish, in which case the gap function acquires opposite signs on the outer and inner pockets.} %AH: Changed text to indicate that the dashed line where the nodes lie is not the unhybridized FS.
	\label{fig:circular_both}
\end{figure*}

Once more, the first step is to diagonalize the Hamiltonian $\mathcal{H}_0+\mathcal{H}_\lambda$ by introducing new pockets $a$ and $b$ exactly as in the case with $\beta=0$. We then rewrite the remaining Hamiltonian in terms of the $a$ and $b$ operators and obtain
\begin{align}\label{eq:H_cbl}
	\mathcal{H}_{\Delta+\beta}' &=\frac{1}{2}\sum_{\mathbf{k}} \left[(\Delta+\beta)a^\dagger_{\mathbf{k}\alpha}a^\dagger_{-\mathbf{k}\beta}+(\Delta-\beta)b^\dagger_{\mathbf{k}\alpha}b^\dagger_{-\mathbf{k}\beta}\right]i\sigma^y_{\alpha\beta}\nonumber\\
	&+\frac{1}{2}\sum_{\mathbf{k}} \Delta y_\mathbf{k} \left[a^\dagger_{\mathbf{k}\alpha}b^\dagger_{-\mathbf{k}\beta}+b^\dagger_{\mathbf{k}\alpha}a^\dagger_{-\mathbf{k}\beta}\right]i\sigma^y_{\alpha\beta}+\mathrm{H.c.}	
\end{align}
 Observe that the coefficient $\beta$ appears only in the intra-pocket terms. We  diagonalize the Hamiltonian $\mathcal{H}_0'+\mathcal{H}_{\Delta+\beta}'$ and again find dispersions of the form $E_\mathbf{k}^\pm= \sqrt{A_\mathbf{k}\pm\sqrt{B_\mathbf{k}}}$, where
\begin{align}
	A_\mathbf{k}=&\xi_\mathbf{k}^2+\Delta^2 (1+y_\mathbf{k}^2)+\lambda^2+\beta^2,\\
	B_\mathbf{k}=&4\left[ (\xi_\mathbf{k}\lambda+\Delta\beta)^2 +\Delta^2 y_\mathbf{k}^2 (\Delta^2+\lambda^2)\right].
\end{align}
The dispersion $E^+_\mathbf{k}$ is fully gapped, but $E^-_\mathbf{k}$ has nodal points at momenta given by
\begin{align}
	\xi_\mathbf{k} &=\lambda\beta/\Delta,\\
	\cos 2\theta_\mathbf{k} &=\pm \frac{\sqrt{\Delta^2+\lambda^2-\beta^2-\lambda^2\beta^2/\Delta^2}}{\alpha\Delta}.
\end{align}
Note that  the nodal points are now shifted from the unhybridized FS. The direction of the shift depends
on the sign of $\beta$. If $\beta>0$ ($\beta<0$) the nodes appear between the original FSs (the ones before hybridization) 
and the outer (inner) hybridized FS. The nodes exist in the parameter range given by
\begin{equation}
	0\leq\Delta^2+\lambda^2-\beta^2-\lambda^2\beta^2/\Delta^2\leq\alpha^2\Delta^2.
\label{ll}
\end{equation}
The lower bound is reached when we keep $\lambda$ fixed and increase $|\beta|$ towards critical $\beta_c = \Delta$.  In this case the nodes merge at the diagonals $k_x=k_y$ or $k_x=-k_y$. The nodes disappear when $|\beta|>\beta_c$ and the intra-pocket gap components in the first line of Eq. (\ref{eq:H_cbl}) have different signs. In the limit of $\lambda\gg \Delta$, the inter-pocket gap component becomes irrelevant since the reconstructed FSs are far apart and in this sense the gap function has opposite signs on the two electron pockets.
 
 The upper boundary in (\ref{ll}) is reached when we set $|\beta| < \Delta$  and increase $\lambda$. In this situation the nodes merge along  the $k_x$ and $k_y$ directions at a critical value of $\lambda$ given by
\begin{equation}
	\lambda_c = \Delta \sqrt{\frac{(\alpha^2-1)\Delta^2+\beta^2}{\Delta^2-\beta^2}}.
\end{equation}
 At $\lambda > \lambda_c$, the nodes disappear and the gap function has the same sign on each electron pocket, as can be clearly seen in the limit of $\lambda\gg\Delta$. Note that $\lambda_c$  grows with $\beta$, i.e., the inter-pocket pairing allows the nodes to exist in a greater range of values of $\lambda$. In this sense the pairing partially protects the nodes from disappearing due to hopping, as long as $|\beta|<\Delta$. The behavior of the nodes when both $\lambda$ and $\beta$ are present is summarized in Fig. \ref{fig:circular_both}.

\section{Elliptical pockets}\label{sec:elliptical}
Now we consider the more realistic case where the electron pockets are elliptical. We will take the dispersions in the form
\begin{align}
	\xi^c_\mathbf{k} &= -\mu +\frac{k_x^2}{2m_1}+\frac{k_y^2}{2m_2},\\
	\xi^d_\mathbf{k} &= -\mu +\frac{k_x^2}{2m_2}+\frac{k_y^2}{2m_1}.
\end{align}
 It is convenient to rewrite the dispersions as $\xi^{c,d}_\mathbf{k}=\xi_\mathbf{k}\pm\delta_\mathbf{k} \cos 2\theta_\mathbf{k}$, where the $+$ sign corresponds to $\xi^c_\mathbf{k}$, $\xi_\mathbf{k}\equiv (\xi_\mathbf{k}^c+\xi_\mathbf{k}^d)/2$, and $\delta_\mathbf{k}\equiv \mathbf{k}^2(m_1^{-1}-m_2^{-1})/4$. Without loss of generality, we will take $\delta_\mathbf{k}$ to be positive.

\subsection{Inter-pocket pairing only ($\beta\neq 0$, $\lambda= 0$)}

\begin{figure*}[htb]
	\centering
		\includegraphics[width=\textwidth]{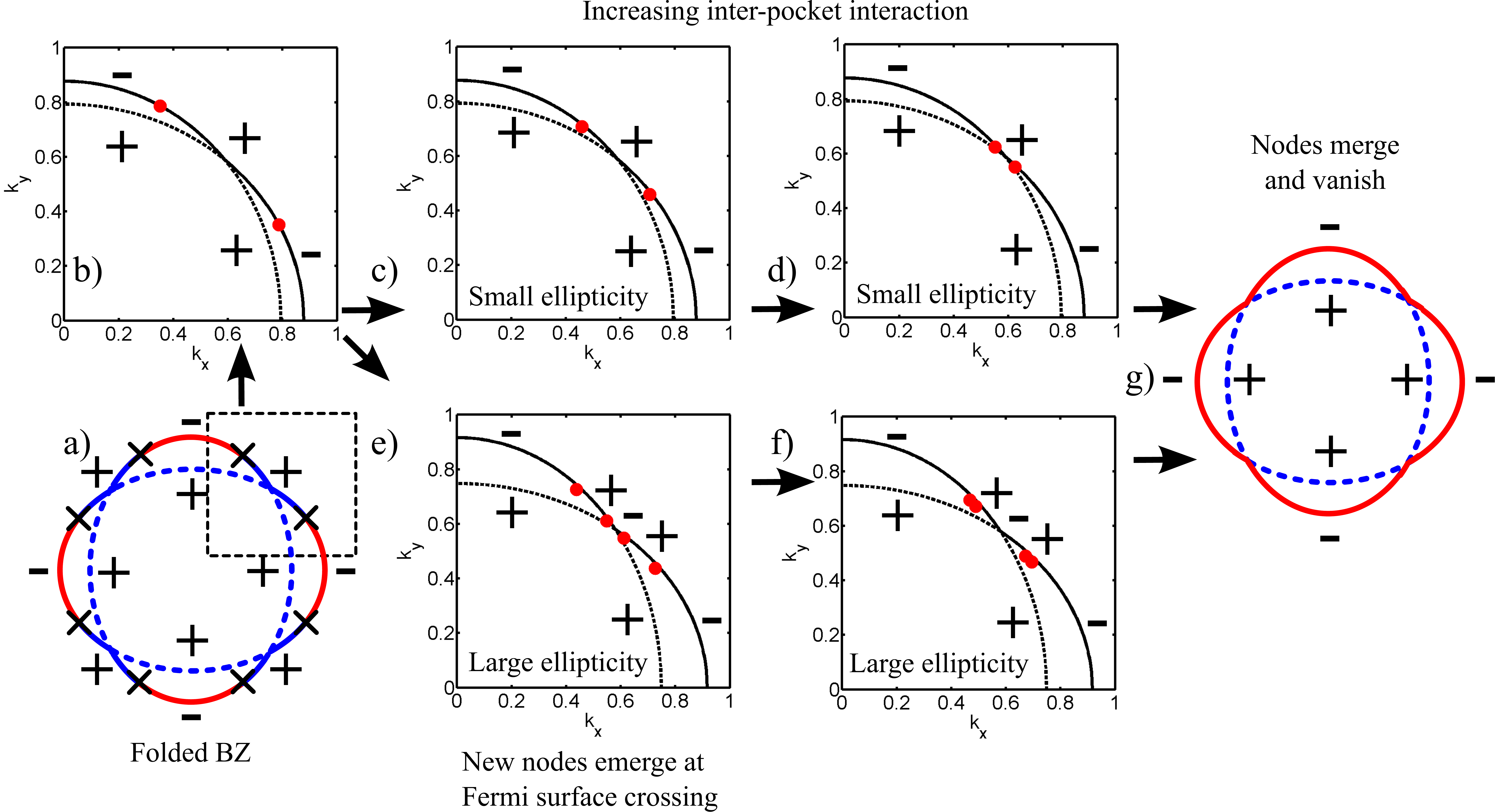}
	\caption{Gap structure for elliptical pockets with inter-pocket pairing. In a) we show the folded electron pockets overlapping. The crosses indicate the position of the nodal points, while the solid and dashed lines indicate opposite signs of the gap function. The evolution of the nodal points as obtained by a numerical calculation is shown in b) through f). First, b) shows the nodal points with no hybridization. If the ellipticity is below a threshold, increasing the inter-pocket pairing simply shifts the nodes toward the diagonal lines $k_x=\pm k_y$ as seen in c) and d). The nodes merge and disappear after reaching the diagonal lines and the gap structure becomes nodeless as shown in g). If this occurs, the gap function has opposite signs on the inner and outer edges of the FSs. Alternatively, if the ellipticity is higher than the threshold, the nodes do not reach the diagonal lines, but instead a node emerges at that symmetry point. As we further increase the inter-pocket pairing the node splits into two nodes which move toward the original nodes, as shown in e) and f). Eventually, the new and old nodes meet and merge, resulting again in the structure shown in g).}
	\label{fig:elliptical_beta}
\end{figure*}

The dispersions after diagonalizing the Hamiltonian are given by $(E_\mathbf{k}^\pm)^2= A_\mathbf{k}\pm\sqrt{B_\mathbf{k}}$, where
\begin{align}
	A_\mathbf{k}=&\frac{1}{2}\Big[(\xi^c_\mathbf{k})^2+(\xi^d_\mathbf{k})^2+2\Delta^2 (1+y_\mathbf{k}^2)+2\beta^2\Big],\\
	B_\mathbf{k}=&\frac{1}{4}\Big[\left((\xi^d_\mathbf{k})^2-(\xi^c_\mathbf{k})^2+4\Delta^2 y_\mathbf{k}\right)^2 \nonumber\\
	&\quad+ 4|\beta|^2\left((\xi^c_\mathbf{k}-\xi^d_\mathbf{k})^2 +4\Delta^2\right) \Big].
\end{align}
Once again, $E^+_\mathbf{k}$ is fully gapped but $E^-_\mathbf{k}$ may contain nodes. Unlike the circular case, the nodes are not located on the original FSs, but at momenta $|\mathbf{k}|$ which are solutions of
\beq\label{eq:k_beta}
\(\delta_\mathbf{k}+\alpha \xi_\mathbf{k}\) \(\xi_\mathbf{k} \delta_\mathbf{k}-\alpha \Delta^2\) +\alpha \beta^2 \delta_\mathbf{k}=0.
\eeq
Let the solutions to this equation be $\xi_\mathbf{k}=\bar{\xi}$ and $\delta_\mathbf{k}=\bar{\delta}$. The angular positions of the nodal points are given by
\begin{equation}
	\cos^2 2\theta_\mathbf{k}= \frac{F(\bar{\xi}, \bar{\delta})}{(\bar{\delta}^2+\alpha^2\Delta^2)^2},
\end{equation}
where
\begin{align}\label{eq:theta_beta}
	F(\bar{\xi}, \bar{\delta}) &=\(\bar{\delta}^2-\alpha^2\Delta^2\) \(\bar{\xi}^2+\beta^2-\Delta^2\) \nonumber\\
		&\quad-4\alpha\Delta^2\bar{\xi}\bar{\delta}.
\end{align}
Note that for each solution to Eq. (\ref{eq:k_beta}) there exist 8 nodal points in the dispersion. One may solve 
  for ${\bar \xi}$ and ${\bar \delta}$ exactly  but the solution is not very illuminating.  It is more useful
   to solve for $\bar{\xi}$ in terms of $\bar{\delta}$ and analyze how the nodal points evolve when we change ${\bar \delta}$.
   Expressing ${\bar \xi}$ in terms of ${\bar \delta}$ we obtain   
\beq
\bar{\xi}=\frac{\bar{\delta} ^2-\alpha ^2 \Delta ^2 \pm\sqrt{\left(\alpha ^2 \Delta ^2+\bar{\delta} ^2\right)^2-4 \alpha ^2 \beta^2 \bar{\delta} ^2}}{2 \alpha  \bar{\delta} }.
\eeq
Substitution of these solutions into Eq. (\ref{eq:theta_beta}) yields
\beq
\cos^2 2\theta_\mathbf{k}=\frac{\bar{\delta} ^2-\alpha ^2 \Delta ^2 \mp\sqrt{\left(\alpha ^2 \Delta ^2+\bar{\delta} ^2\right)^2-4 \alpha ^2 \beta^2 \bar{\delta} ^2}}{2 \alpha^2  \bar{\delta}^2 }
\label{a_3}
\eeq

 Analyzing (\ref{a_3}) we find new interesting physics. Namely, depending on the parameters, there may be 0, 8, or 16 nodal points in the dispersion. When
  $\bar{\delta}<\alpha\Delta$,  there are either 8 or zero nodes, as one can immediately verify.  At small $\beta$, there are 8 nodes. 
      As $|\beta|$  increases, the nodes move toward the diagonals $k_x=\pm k_y$, like in the circular case. At $|\beta|=\Delta$, pairs of nodes merge,
        and for $|\beta|>\Delta$ they disappear.   The outcome of the disappearance of the nodes is the effective $s^{+-}$ superconducting state 
          with different signs of the gap on the inner and outer electron pockets, see Fig. \ref{fig:elliptical_beta}.

 When $\bar{\delta}>\alpha\Delta$, the evolution of the nodes is more interesting. At small $\beta$, there are again 8 nodes. As $|\beta|$ increases, the nodes shift towards diagonals but they do not reach $k_x=\pm k_y$ at $|\beta| =\Delta$. Instead, at this $\beta$, a new quadratic node appears in in each quadrant at the point where zone diagonals intersect the original FS.  At $|\beta| > \Delta$, each quadratic node splits into two, one to the right and one to the left of a diagonal, and each new node moves toward the already existing nodes (see Fig. \ref{fig:elliptical_beta}).  Thus, there is a total of 16 nodal points. 
  As  $|\beta|$ continues increasing,  the old and new nodes merge
  at a critical value $|\beta|=\beta_c$ given by
  \beq
  \beta_c=\left(\alpha ^2 \Delta ^2+\bar{\delta} ^2\right)/ (2|\alpha|  \bar{\delta}).
   \eeq
   The nodes disappear when $|\beta|$ exceeds this critical value and the end result of the evolution of the nodes is the same minus-plus gap on the inner and outer electron pockets.

We verified this behavior numerically (Fig. \ref{fig:elliptical_beta} actually shows the results of numerical calculations).
 In all numerical examples here and below we have set $\mu=10\Delta$ and $\alpha=-1.5$.
Note in passing that while the relationships presented in this analysis are exact, one must keep in mind that in general $\bar{\delta}$ by itself depends on $\beta$. 

\subsection{Inter-pocket hopping only ($\lambda\neq 0$, $\beta=0$)}

\begin{figure*}[htb]
	\centering
		\includegraphics[width=0.8\textwidth]{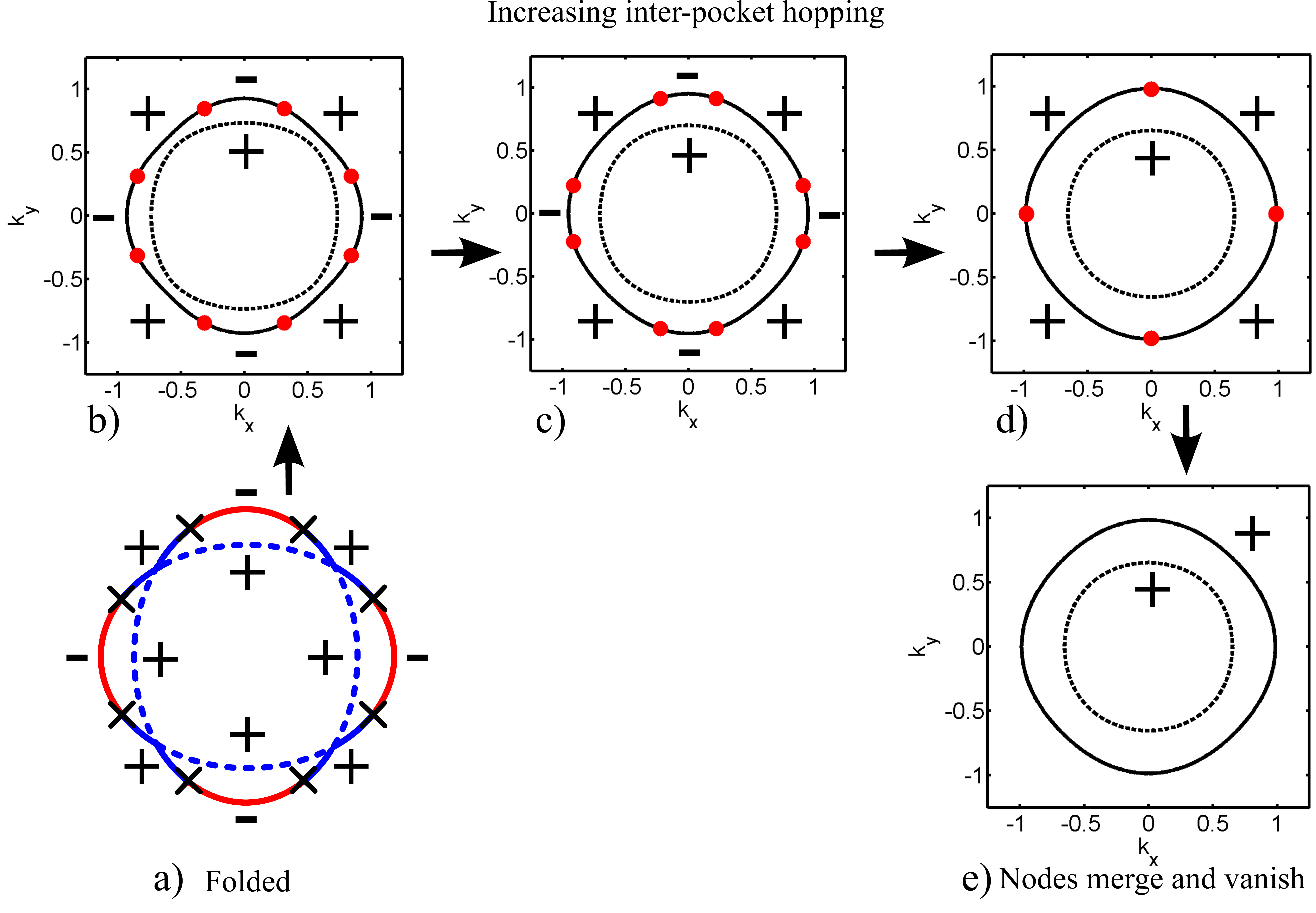}
	\caption{Gap structure for elliptical pockets with inter-pocket hopping. In a) we show the gap structure in the absence of hybridization. Inter-pocket hopping reconstructs the FSs as shown in b). As the hopping parameter increases, the inner FS shrinks and the outer one becomes larger. In the superconducting state, the nodal points (represented as crosses) lie near, but not exactly on FSs and shift toward the $k_x$ and $k_y$ axis as the hopping increases, as shown in b) and c). At a critical value of the hopping the nodes meet and merge in pairs as in d). Finally, for greater values of the parameter they vanish and the sign of the gap function becomes the same on both FSs. Subfigures b) through e) are the result of numerical calculations.}
	
\label{fig:elliptical_lambda}
\end{figure*}

This case has been studied before~\cite{Khodas2012b}  and we present it here for completeness, using a somewhat different computation scheme.
 The first step is to diagonalize the Hamiltonian $\mathcal{H}_0+\mathcal{H}_\lambda$ by introducing new operators $a$ and $b$ such that
\begin{align}
	c_{\mathbf{k}\alpha} &=u_\mathbf{k}c_{\mathbf{k}\alpha}+v_\mathbf{k}b_{\mathbf{k}\alpha}, \nonumber\\
	d_{\mathbf{k}\alpha} &=-v_\mathbf{k}a_{\mathbf{k}\alpha}+u_\mathbf{k}b_{\mathbf{k}\alpha},
\end{align}
where $u_\mathbf{k}=\cos \psi$ and $v_\mathbf{k}=-\sin \psi$, and this angle $\psi$ satisfies $\cos 2\psi=(\xi_\mathbf{k}^c-\xi_\mathbf{k}^d)/\sqrt{(\xi_\mathbf{k}^c-\xi_\mathbf{k}^d)^2+4\lambda^2}$ and $\sin 2\psi=2\lambda/\sqrt{(\xi_\mathbf{k}^c-\xi_\mathbf{k}^d)^2+4\lambda^2}$. The new Hamiltonian is
\begin{equation}
	\mathcal{H}_0' =\sum_{\mathbf{k}} \left[\xi^a_\mathbf{k}a^\dagger_{\mathbf{k}\alpha}a_{\mathbf{k}\alpha}+\xi^b_\mathbf{k}b^\dagger_{\mathbf{k}\alpha}b_{\mathbf{k}\alpha}\right],
\end{equation}
where $\xi^{a,b}_\mathbf{k}=\frac{1}{2}(\xi^c_\mathbf{k}+\xi^d_\mathbf{k})\pm\sqrt{\lambda^2+(\xi_\mathbf{k}^c-\xi_\mathbf{k}^d)^2/4}$. These new dispersions define the reconstructed FSs  shown in Fig. \ref{fig:elliptical_lambda}. As the hopping parameter $\lambda$ increases, the outer FS (associated with $b$ fermions) becomes larger, while the inner FS (associated with $a$ fermions) shrinks.

Now we consider the superconducting state. Rewriting the pairing part of the Hamiltonian in terms of the new operators we find
\begin{align}
	\mathcal{H}_{\Delta}' &=\frac{1}{2}\sum_{\mathbf{k}} \Big[\Delta(1-f_\mathbf{k} y_\mathbf{k})a^\dagger_{\mathbf{k}\alpha}a^\dagger_{-\mathbf{k}\beta} \Big.\nonumber\\
	&\quad\quad\quad+\Big.\Delta(1+f_\mathbf{k} y_\mathbf{k})b^\dagger_{\mathbf{k}\alpha}b^\dagger_{-\mathbf{k}\beta}\Big]i\sigma^y_{\alpha\beta}\\
	&-\frac{1}{2}\sum_{\mathbf{k}} \Delta g_\mathbf{k}y_\mathbf{k} \left[c^\dagger_{\mathbf{k}\alpha}d^\dagger_{-\mathbf{k}\beta}+d^\dagger_{\mathbf{k}\alpha}c^\dagger_{-\mathbf{k}\beta}\right]i\sigma^y_{\alpha\beta}+\mathrm{H.c.},	\nonumber
\end{align}
where $f_\mathbf{k}\equiv \cos 2\psi$ and $g_\mathbf{k}\equiv -\sin 2\psi$. The diagonalization of this Hamiltonian again yields dispersions in the superconducting state in
 the form $(E_\mathbf{k}^\pm)^2= A_\mathbf{k}\pm\sqrt{B_\mathbf{k}}$. In this particular case
\begin{align}
	A_\mathbf{k}=&\frac{1}{2}\Big[(\xi^a_\mathbf{k})^2+(\xi^b_\mathbf{k})^2+2\Delta^2 (1+y_\mathbf{k}^2)\Big],\\
	B_\mathbf{k}=&\frac{1}{4}\Big[\left((\xi^b_\mathbf{k})^2-(\xi^a_\mathbf{k})^2+4\Delta^2 f_\mathbf{k}y_\mathbf{k}\right)^2 \nonumber\\
	&\quad+ 4\Delta^2 y^2_\mathbf{k}g^2_\mathbf{k}\left((\xi^a_\mathbf{k}-\xi^b_\mathbf{k})^2 +4\Delta^2\right) \Big].
\end{align}

As usual, the dispersion $E^+_\mathbf{k}$ is nodeless, but $E^-_\mathbf{k}$ has nodes at momenta which are the solutions of the equation
\beq\label{eq:k_lambda}
\(\delta_\mathbf{k}+\alpha \xi_\mathbf{k}\) \(\xi_\mathbf{k} \delta_\mathbf{k}-\alpha \Delta^2\) -\alpha \lambda^2 \delta_\mathbf{k}=0.
\eeq

Each solution to this equation defines a pair ($\bar{\xi}$, $\bar{\delta}$) and determines the radial position of the nodal point. The angular position is then given by
\begin{equation}\label{eq:theta_lambda}
	\cos^2 2\theta_\mathbf{k}= \frac{F(\bar{\xi}, \bar{\delta})}{(\bar{\delta}^2+\alpha^2\Delta^2)^2},
\end{equation}
where
\begin{align}
	F(\bar{\xi}, \bar{\delta}) &=\(\bar{\delta}^2-\alpha^2\Delta^2\) \(\bar{\xi}^2-\lambda^2-\Delta^2\) \nonumber\\
		&\quad-4\alpha\Delta^2\bar{\xi}\bar{\delta}.
\end{align}

Like before, we solve for $\bar{\xi}$ in terms of $\bar{\delta}$. The solution is
\beq
\bar{\xi}=\frac{-\bar{\delta} ^2+\alpha ^2 \Delta ^2 -\sqrt{\left(\alpha ^2 \Delta ^2+\bar{\delta} ^2\right)^2+4 \alpha ^2 \lambda^2 \bar{\delta} ^2}}{2 \alpha  \bar{\delta} }.
\eeq
When we substitute this solution into Eq. (\ref{eq:theta_lambda}) we find that the angular position of the nodes is given by
\beq
\cos^2 2\theta_\mathbf{k}=\frac{\bar{\delta} ^2-\alpha ^2 \Delta ^2 +\sqrt{\left(\alpha ^2 \Delta ^2+\bar{\delta} ^2\right)^2+4 \alpha ^2 \lambda^2 \bar{\delta} ^2}}{2 \alpha^2  \bar{\delta}^2 }.
\eeq

The analysis of these equations shows that nodal points appear in a set of 8 and that they are not located on the FSs of the normal state, although our numerical calculations show that they remain very close to it. The location of the nodes with respect to the original FSs varies depending on the sign of $\alpha$. If $\alpha>0$, the nodes are located inside of both unhybridized FSs but outside of the smaller FS. Instead, if $\alpha<0$ the nodes are outside the unhybridized FSs but inside the larger reconstructed FS.

In both cases, the behavior is qualitatively the same as in the limiting case of circular pockets and is summarized in Fig. \ref{fig:elliptical_lambda}, where we show the result of numerical calculations. Increasing $\lambda$ shifts the nodes toward the $k_x$ and $k_y$ axes. The critical value of $\lambda$ that causes the nodes to merge along these directions is enhanced by the ellipticity and is given by $\lambda_c=\sqrt{\(\Delta^2+\bar{\delta}^2\) \(\alpha^2-1\)}$. At any larger $\lambda$ the nodes disappear.
In this sense, the eccentricity of the pockets tries to prevent the disappearance of the nodes.

Regarding the gap structure, we note that for large $\lambda$ the inter-pocket pairing term is irrelevant as the reconstructed FSs are far apart from each other. In this limit, the gap on the reconstructed pockets is given by $\Delta(1\pm f_\mathbf{k}y_\mathbf{k})$, %AH: The factor of \Delta was missing.
 where $|f_\mathbf{k}y_\mathbf{k}|\ll 1$. Thus, the phase of the gap function is equal and uniform over the the reconstructed FSs.  
%AC
 At smaller $\lambda$, the phase of the gap along the FSs cannot be determined as the pairing involves both intra-pocket and inter-pocket terms. 
Judging from the large $\lambda$ limit, it seems natural to define the gap with equal sign on both FSs also at intermediate $\lambda$, see Fig. \ref{fig:elliptical_lambda}.

\subsection{Inter-pocket pairing and hopping ($\beta\neq0$, $\lambda\neq 0$)}

\begin{figure*}[htb]
	\centering
		\includegraphics[width=0.8\textwidth]{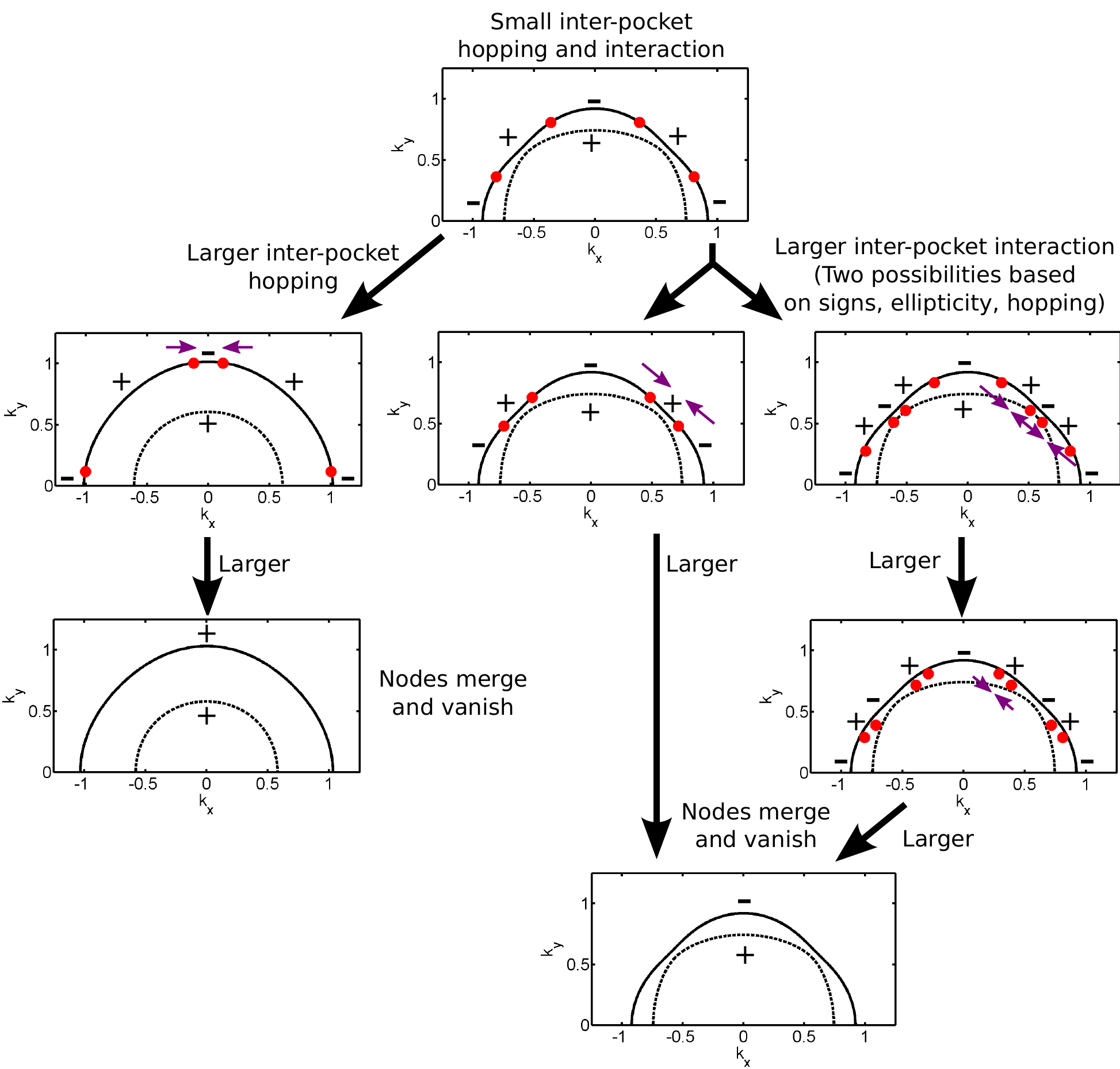}
	\caption{Gap structure for elliptical pockets with inter-pocket hopping and pairing. This case incorporates all the features seen in Figs. \ref{fig:elliptical_beta} and \ref{fig:elliptical_lambda}. The main difference is that the condition for additional nodes does not depend solely on the ellipticity, but also on the hopping parameter and the sign of the inter-pocket pairing.}
	\label{fig:elliptical_all}
\end{figure*}

This case is the most generic one. Like before, we switch to the reconstructed $a$ and $b$ pockets after hybridization. The pairing terms in the Hamiltonian can be rewritten in terms of $a$ and $b$  operators and take the form
\begin{align}
	\mathcal{H}_{\Delta+\beta}' &=\frac{1}{2}\sum_{\mathbf{k}} \left[\Delta_{aa}(\mathbf{k})a^\dagger_{\mathbf{k}\alpha}a^\dagger_{-\mathbf{k}\beta}+\Delta_{bb}(\mathbf{k})b^\dagger_{\mathbf{k}\alpha}b^\dagger_{-\mathbf{k}\beta}\right]i\sigma^y_{\alpha\beta}\nonumber\\
	&+\frac{1}{2}\sum_{\mathbf{k}} \Delta_{ab}(\mathbf{k}) \left[a^\dagger_{\mathbf{k}\alpha}b^\dagger_{-\mathbf{k}\beta}+b^\dagger_{\mathbf{k}\alpha}a^\dagger_{-\mathbf{k}\beta}\right]i\sigma^y_{\alpha\beta}+\mathrm{H.c.},	
\end{align}
where
\begin{align}
	\Delta_{aa}(\mathbf{k}) &=\Delta(1-y_\mathbf{k}f_\mathbf{k}) -\beta g_\mathbf{k},\\
	\Delta_{bb}(\mathbf{k}) &=\Delta(1+y_\mathbf{k}f_\mathbf{k}) +\beta g_\mathbf{k},\\
	\Delta_{ab}(\mathbf{k}) &=-\Delta y_\mathbf{k}g_\mathbf{k} +\beta f_\mathbf{k}.
\end{align}

The dispersions are given by $(E_\mathbf{k}^\pm)^2= A_\mathbf{k}\pm\sqrt{B_\mathbf{k}}$,
where
\begin{align}
	A_\mathbf{k}=&\frac{1}{2}\Big[\(\xi^a_\mathbf{k}\)^2+\(\xi^b_\mathbf{k}\)^2+2\Delta^2 \(1+y_\mathbf{k}^2\) +2\beta^2\Big],\\
	B_\mathbf{k}=&\frac{1}{4}\Big[\left(\(\xi^b_\mathbf{k}\)^2-\(\xi^a_\mathbf{k}\)^2+4\Delta^2 f_\mathbf{k}y_\mathbf{k}\right)^2 \nonumber\\
	&\quad+ 4\Delta^2 y^2_\mathbf{k}g^2_\mathbf{k}\left((\xi^a_\mathbf{k}-\xi^b_\mathbf{k})^2 +4\Delta^2\right) \Big]\nonumber\\
	-&2\beta\Delta  g_\mathbf{k} \(\xi_\mathbf{k}^a-\xi_\mathbf{k}^b\) \[ \xi^a_\mathbf{k} +\xi^b_\mathbf{k} +\(\xi^a_\mathbf{k}-\xi^b_\mathbf{k}\) f_\mathbf{k} y_\mathbf{k}\]\nonumber\\
	+&\beta^2 \[4\Delta^2+\(\xi^a_\mathbf{k}-\xi^b_\mathbf{k}\)^2 f^2_\mathbf{k}\].	
\end{align}

Once more, we search for nodes in the dispersion $E^-_\mathbf{k}$. The radial position of the nodes is determined by the condition
\begin{align}\label{eq:k_both}
\left(\delta_\mathbf{k}  \xi_\mathbf{k} -\alpha  \Delta^2\right)(\delta_\mathbf{k}+\alpha  \xi_\mathbf{k}) &\nonumber\\
+\frac{\beta  \lambda}{\Delta}  \(\alpha^2 \Delta^2-\delta^2_\mathbf{k}\)
+\alpha \(\beta^2-\lambda^2\)\delta_\mathbf{k} & =0.
\end{align}
The solutions to this equation $(\bar{\xi},\bar{\delta})$ are needed to find the angular position of the nodes:
\begin{equation}\label{eq:theta_both}
	\cos^2 2\theta_\mathbf{k}= \frac{F(\bar{\xi}, \bar{\delta})}{(\bar{\delta}^2+\alpha^2\Delta^2)^2},
\end{equation}
where
\begin{align}
	F(\bar{\xi}, \bar{\delta}) &=\(\bar{\delta}^2-\alpha^2\Delta^2\) \(\bar{\xi}^2+\beta^2-\lambda^2-\Delta^2\) \nonumber\\
		&\quad+4\alpha\Delta\bar{\delta}\(\beta\lambda-\Delta\bar{\xi}\).
\end{align}

\begin{widetext}
Solving for $\bar{\xi}$ in terms of $\bar{\delta}$, we find that the solutions to Eq. (\ref{eq:k_both}) are
\begin{equation}
\bar{\xi} =\frac{1}{2 \alpha  \bar{\delta}} \Bigg\{-\bar{\delta} ^2+\alpha ^2 \Delta ^2 \pm\Bigg[\left(\alpha ^2 \Delta ^2+\bar{\delta} ^2\right)^2+4\alpha\bar{\delta}\bigg(\(\bar{\delta}^2-\alpha^2\Delta^2\) \frac{\lambda\beta}{\Delta}+ \alpha\bar{\delta}\(\lambda^2-\beta^2\)\bigg)\Bigg]^{1/2}\Bigg\}.
\end{equation}
The angular location of the nodes for these solutions is given by
\begin{equation}
\cos^2 2\theta_\mathbf{k} =\frac{1}{2 \alpha^2  \bar{\delta}^2} \Bigg\{\bar{\delta} ^2-\alpha ^2 \Delta ^2 +2\alpha\delta\beta\lambda/\Delta\mp\Bigg[\left(\alpha ^2 \Delta ^2+\bar{\delta} ^2\right)^2+4\alpha\bar{\delta}\bigg(\(\bar{\delta}^2-\alpha^2\Delta^2\) \frac{\lambda\beta}{\Delta}+ \alpha\bar{\delta}\(\lambda^2-\beta^2\)\bigg)\Bigg]^{1/2}\Bigg\}.
\end{equation}
\end{widetext}

In this general case the interplay of the different parameters is considerably more complicated that in the previous limiting cases, but it does not produce any new features. We find that in general the position of the nodal points depends on the signs of both $\beta$ and $\alpha$, not only their magnitudes. The dependence on the sign of $\beta$ comes from bilinear terms of the form $\beta\lambda$, while the dependence on the sign of $\alpha$ is a consequence of the ellipticity of the pockets.

The general behavior of the nodal points is summarized in Fig. \ref{fig:elliptical_all}. We found by numerical analysis that in general increasing $\lambda$ tends to shift the nodes toward the $k_x$ and $k_y$ axis as usual, where they merge and disappear at a critical value of $\lambda$. In this  case, the sign of the order parameter is the same on both FSs. Increasing $|\beta|$ instead shifts the nodes toward the diagonals $k_x=\pm k_y$. At $|\beta|=\Delta$ there are two possibilities, as seen in the limiting case of $\lambda=0$. The first is that the eight nodal points merge in pairs at the diagonal lines, disappearing for $|\beta|>\Delta$. 
%AC
 This happens at small ${\bar \delta}$, i.e., for small eccentricity. 
  The second is that the original nodal points do not reach the diagonal lines at this value of $\beta$, but instead four new nodes appear at those lines. Increasing $|\beta|$ further causes these four new nodal points to split into pairs, and moves the old and new nodes toward each other. %AH: Changed "split into either" to "split into pairs".
	At a threshold value of $\beta$ they merge and then disappear. This second scenario, with 16 nodal points at intermediate $\beta$, is realized at larger ${\bar \delta}$, i.e., at larger eccentricity. In either case, the merging of nodes caused by large $\beta$ means that the gap function has opposite signs on the electron pockets.

One important difference with the case of $\lambda=0$ is that the condition for developing additional nodal points is more complicated since the sign of $\beta$ and the value of $\lambda$ also play a role. It is clear from the equations that the additional nodes are more likely to develop for $\beta<0$, for $\delta>\alpha\Delta$ and for small $\lambda$. The exact conditions when additional nodes appear are given by rather involved formulas and we refrain from 
 presenting them.

\section{Conclusions}\label{sec:conclusions}
In this paper we have investigated the effect of hybridization of the two electron pockets on the gap structure in FeSCs. We considered the case when the dominant pairing interaction is between hole and electron pockets and it yields an $s^{+-}$ gap with accidental nodes on the electron pockets.
Our goal was to understand how accidental nodes move once we include the hybridization.
 We argued that for an $s^{+-}$ superconductor hybridization gives rise to two effects --  hopping between electron FSs and the appearance of an additional pairing term which describes inter-pocket pairing.
  Each of these two effects shifts the position of the nodes and at large enough hybridization the nodes eventually disappear. However, the evolution of the nodes and the gap structure of the resulting no-nodal state is different, depending on whether the inter-pocket hopping or the inter-pocket interaction is stronger.
 In the first case, the resulting state has the same sign of the gap on both reconstructed FSs.  In the second case, there is a sign change of the superconducting gap  between the inner and outer FSs.
  %(in the absence of inter-pocket hopping), or between the new FSs and the original FS (in the presence of inter-pocket hopping). 
  We also found that the evolution of the nodes with increasing inter-pocket pairing interaction is rather non-trivial, and
   in the intermediate regime the number of nodal points may increase from 8 to 16.
 We also found that the eccentricity of the pockets enlarges these critical values, partially protecting the nodal points from disappearing. The bottom line of this analysis is that  strong hybridization lifts accidental nodes, but the  resulting superconducting state may be highly non-trivial, particularly when the dominant effect of hybridization is the emergence of inter-pocket pairing potential.

We acknowledge useful conversations with M. Khodas, P. Coleman, R. Fernandes, K. Haule, A. Kamenev, and G. Kotliar. %AH: Added M. Khodas, since we discussed this with him when we were starting.
The work is supported by the DOE grant DE-FG02-ER46900.


\begin{thebibliography}{10}
\bibitem{hosono}
Y. Kamihara, T. Watanabe, M. Hirano, H.
Hosono, J. Am. Chem. Soc. \textbf{130}, 3296(2008).
\bibitem{BaFeAsP} M. Yamashita, Y. Senshu, T. Shibauchi, S. Kasahara, K. Hashimoto, D. Watanabe, H. Ikeda, T. Terashima, I. Vekhter, A. B. Vorontsov, and Y. Matsuda, Phys.Rev.B 84, 060507(R) (2011).
\bibitem{LaOFeP} A. I. Coldea, J. D. Fletcher, A. Carrington, J. G. Analytis, A. F. Bangura, J.-H. Chu, A. S. Erickson, I. R. Fisher, N. E. Hussey, and R. D. McDonald, Phys.Rev.Lett. 101, 216402 (2008).
\bibitem{LiFeP} S. Kasahara, K. Hashimoto, H. Ikeda, T. Terashima, Y. Matsuda, and T. Shibauchi, Phys. Rev. B 85, 060503(R) (2012); K. Hashimoto, S. Kasahara, R. Katsumata, Y. Mizukami, M. Yamashita, H. Ikeda, T. Terashima, A. Carrington, Y. Matsuda, and T. Shibauchi, Phys. Rev. Lett. 108, 047003 (2012).
\bibitem{review}
S. Graser, T. A. Maier, P. J. Hirshfeld, D. J. Scalapino, New
J. Phys. \textbf{11}, 025016 (2009); A. F. Kemper, T. A. Maier, S. Graser, H.-P. Cheng, P. J. Hirschfeld, and D. J. Scalapino, New J. Phys. {\bf 12},  073030 (2010);
 K. Kuroki, H. Usui, S. Onari, R. Arita, and H. Aoki, Phys. Rev. B
79, 224511 (2009); A. V. Chubukov,
Physica C \textbf{469}, 640(2009); I.I. Mazin and J. Schmalian,  Physica C,
\textbf{469}, 614 (2009); D.N. Basov and A.V. Chubukov, Nature Physics {\bf 7}, 241 (2011); P.J. Hirschfeld, M.M. Korshunov, and I.I. Mazin, Reports
on Progress in Physics 74, 124508 (2011); A.V. Chubukov, Annual Review of Condensed Matter
Physics 3, 57 (2012); A.V. Chubukov,  "Itinerant electron scenario" in
Springer Series in Materials Science, {\bf 211} 255 (2015).
\bibitem{Kemper2010} A. F. Kemper, T. A. Maier, S. Graser, H.-P. Cheng, P. J. Hirschfeld, and D. J. Scalapino, New J. Phys. \textbf{12}, 073030 (2010).
\bibitem{vavilov} A.V. Chubukov, M.G. Vavilov, A.B. Vorontsov
Phys. Rev. B \textbf{80}, 140515(R) (2009); T. A. Maier, S. Graser, D. J. Scalapino, and P. J. Hirschfeld, Phys. Rev. B {\bf 79}, 224510 (2009).
\bibitem{Maiti2011}  S. Maiti, M. M. Korshunov, T. A. Maier, P. J. Hirschfeld, and A. V. Chubukov, Phys. Rev. B \textbf{84}, 224505 (2011).
\bibitem{hybrid} K. Suzuki, H. Usui, and K. Kuroki, J. Phys. Soc. Jpn. 80
013710 (2011);  T. Saito, S. Onari, H. Kontani,
 Phys. Rev. B {\bf 83}, 140512(R) (2011); Chia-Hui Lin, Tom Berlijn,  Limin Wang, Chi-Cheng Lee, Wei-Guo Yin, and Wei Ku, Phys. Rev. Lett.,  107, 257001 (2011);
 J-P Hu and N. Hao, Phys. Rev. X 2, 021009 (2012); M.I. Calderon, B. Valenzuela, and E. Bascones, Phys. Rev. {\bf 80}, 094531 (2009); 
 A. Moreo, M. Daghoger, J.A. Riera, and E. Dagotto, Phys. Rev. B {\bf 79}, 13452 (2009); M.
 Daghoger, A. Nicholson, A. Moreo, and E. Dagotto, Phys. Rev. B {\bf 81}, 014511 (2010); T. Miyake, K. Nakamura, R. Arita, M. Imada, J. Phys.
Soc. Jpn. {\bf 79} 044705 (2010).
\bibitem{Khodas2012b} M. Khodas and A. V. Chubukov, Phys. Rev. B \textbf{86}, 144519 (2012).
\bibitem{Kotliar2014} Z. P. Yin, K. Haule, and G. Kotliar, Nat Phys \textbf{10}, 845 (2014).
\bibitem{ilya_li}
F. Ahn, I. Eremin, J. Knolle, V.B. Zabolotnyy, S.V. Borisenko, B. Büchner, and A.V. Chubukov,
 Phys. Rev. B 89, 144513 (2014).
 \bibitem{oskar} V. Cvetkovic and O. Vafek, Phys. Rev. B 88, 134510 (2013). 
 \bibitem{mazin_i} I. I. Mazin, Phys. Rev. B 84, 024529 (2011).
\bibitem{sadovskii} I.A. Nekrasov, Z.V. Pchelkina, M.V. Sadovskii,
JETP Letters, {\bf 88}, 144 (2008);
Y. Su, P. Link, A. Schneidewind, Th. Wolf, P. Adelmann, Y. Xiao, M. Meven, R. Mittal, M. Rotter, D. Johrendt, Th. Brueckel, and M. Loewenhaupt,
Phys. Rev. B {\bf 79}, 064504 (2009);
J. Guo, S. Jin, G. Wang, Sh. Wang, K. Zhu, T. Zhou, M. He, and X. Chen,
Phys. Rev. B {\bf 82}, 180520(R) (2010).
\bibitem{Coldea} A. I. Coldea,
Philos. Trans. R. Soc. A 368, 3503 (2010); A. Carrington, A. I. Coldea, J. D. Fletcher, N. E. Hussey, C. M.
J. Andrew, A. F. Bangura, J. G. Analytis, J.-H. Chu, A. S. Erickson, I. R. Fisher, and R. D. McDonald, Physica C 469, 459 (2009).
\bibitem{Khodas2012a} M. Khodas and A. V. Chubukov, Phys. Rev. Lett. \textbf{108}, 247003 (2012).

\end{thebibliography}
\end{document}